%% file: template.tex
\documentclass{appolb}
\usepackage{ifthen} 
\newboolean{pdflatex}
\setboolean{pdflatex}{true}
\newboolean{articletitles}
\setboolean{articletitles}{true} 

\newboolean{uprightparticles}
\setboolean{uprightparticles}{false} 
\newboolean{inbibliography}
\setboolean{inbibliography}{false}

\usepackage{graphicx}
\usepackage{mciteplus}
\usepackage{longtable}
\usepackage{hyperref}
\input{preamble}

\begin{document}
\title{Quarkonia production in ultra-peripheral PbPb collisions at LHCb%
\thanks{Presented at "Diffraction and Low-$x$ 2022", Corigliano Calabro (Italy), September 24-30, 2022}%
}
\author{Xiaolin Wang on behalf of the LHCb collaboration 
\address{Guangdong Provincial Key Laboratory of Nuclear Science, Guangdong-Hong Kong Joint Laboratory of Quantum Matter, Institute of Quantum Matter, South China Normal University, Guangzhou, China}
\\[3mm]
}
\maketitle
\begin{abstract}
Measurements of coherent charmonium production cross sections together with their ratio in ultra-peripheral PbPb collisions are studied at a nucleon-nucleon centre-of-mass energy of $5.02\,\mathrm{TeV}$,
the differential cross-sections are measured as a function of rapidity and transverse momentum, separately.
The photo-production of \jpsi mesons at low transverse momentum is studied in peripheral PbPb collisions, which confirms coherent \jpsi production in hadronic collisions.
These latest results significantly improve previous measurements and are compared with some theoretical predictions.
\end{abstract}
  
\section{Introduction}
The $\lhcb$ detector is a single-arm forward spectrometer covering the \mbox{pseudorapidity} range $2<\eta <5$~\cite{LHCb-DP-2008-001, LHCb-DP-2014-002}, 
which has great performance with precise vertex reconstruction, high particle momentum resolution, and excellent particle identification capability. 
Measurements of quarkonia production in (ultra-)peripheral heavy-ion collisions play an important role in studying the photon-nucleus interaction, the mechanism of vector meson production, and the partonic structure of nuclei. Meanwhile, coherent photo-production would provide an excellent laboratory to study the nuclear shadowing effects and the initial state of heavy ion collisions at small-$x$ at the \lhc~\cite{Jones:2015nna}. 

\section{Charmonia production in ultra-peripheral PbPb collisions at LHCb~\cite{LHCb:2022ahs} }
Ultra-peripheral collisions(UPCs) occur when the impact parameter is larger than the sum of the radii of two incoming nuclei~\cite{Bertulani:2005ru}. 
The two ions interact via their cloud of semi-real photons, and photon-nuclear interactions dominate. 
In UPCs, \jpsi and \psitwos mesons are produced from the colorless reaction of a photon from one nucleus and a Pomeron from the other.
If the photon interacts with a Pomeron emitted by the whole nucleus, there would be no nucleus break-up, and this is called coherent production, which is the process we are going to study.

\begin{figure}[htbp]
    \centering
    \hfil
    \begin{center}
     \begin{minipage}[t]{0.50\linewidth}  
  \includegraphics[width=\linewidth, page={1}]{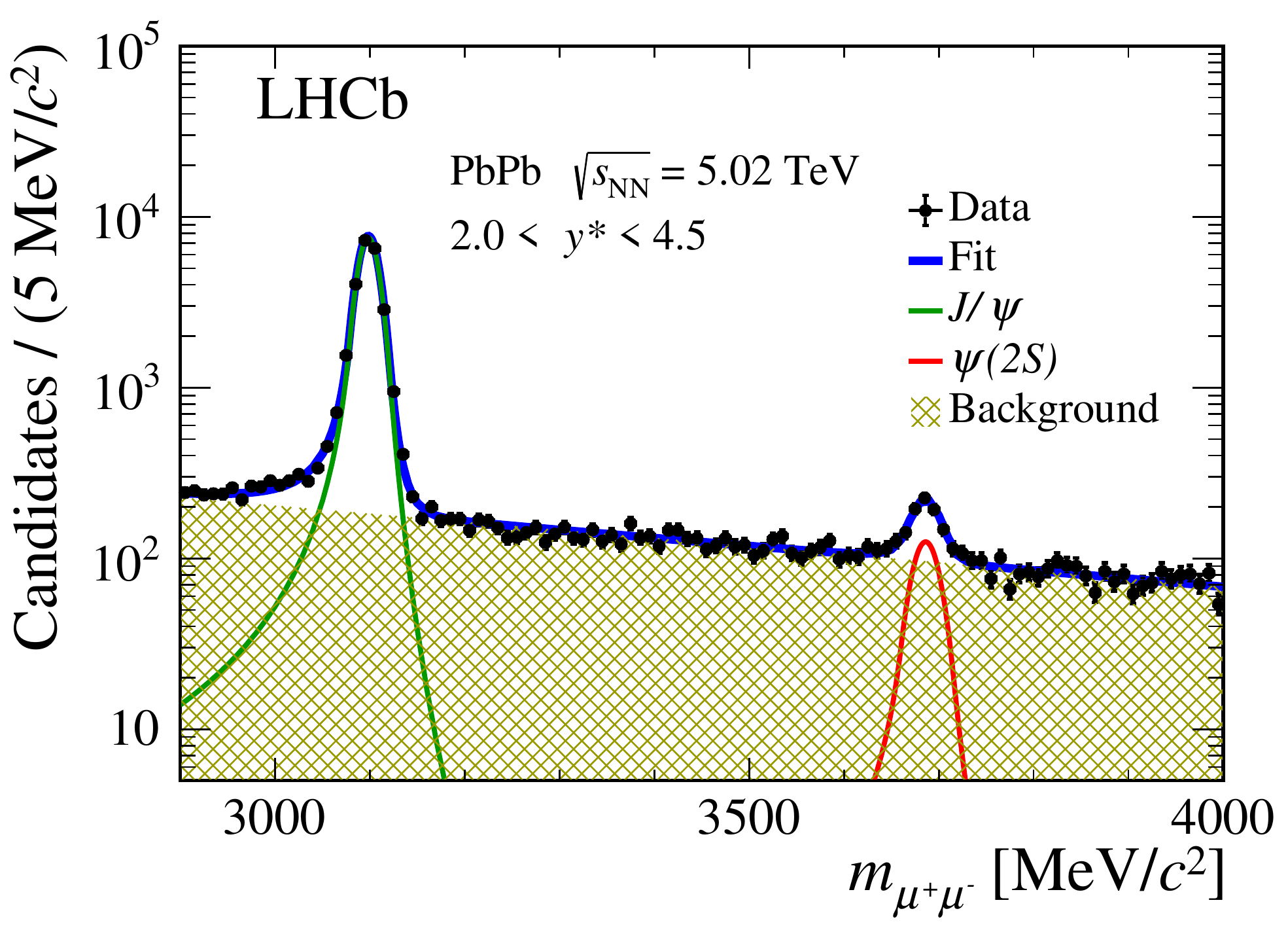}      
    \end{minipage}     
    \end{center}
    \begin{minipage}[t]{0.49\linewidth}
        \centering
    \includegraphics[width=\linewidth, page = {1}]{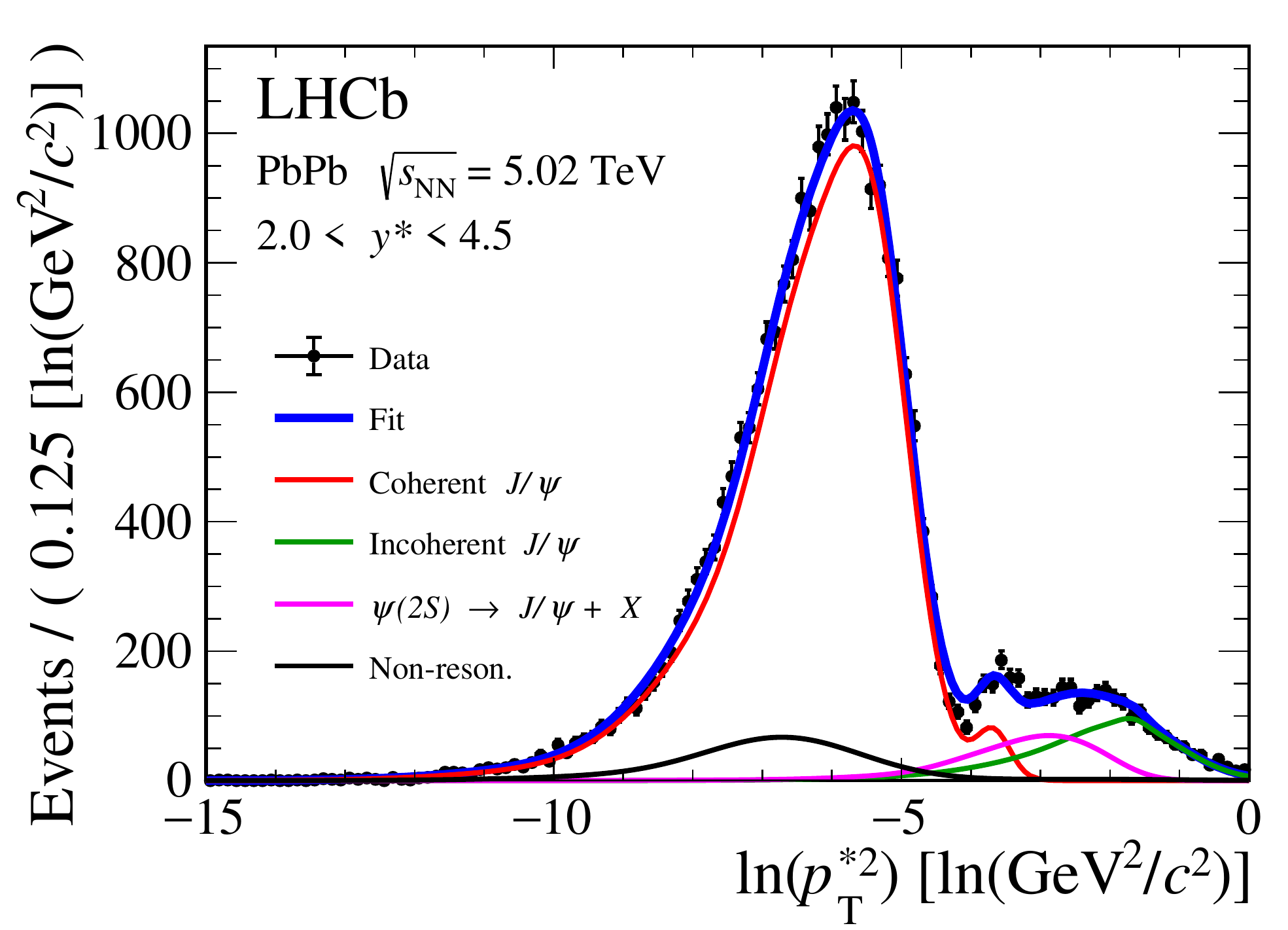}
        \put(-40,100){\jpsi}
    \end{minipage}
    \begin{minipage}[t]{0.49\linewidth}
        \centering
    \includegraphics[width=\linewidth, page = {1}]{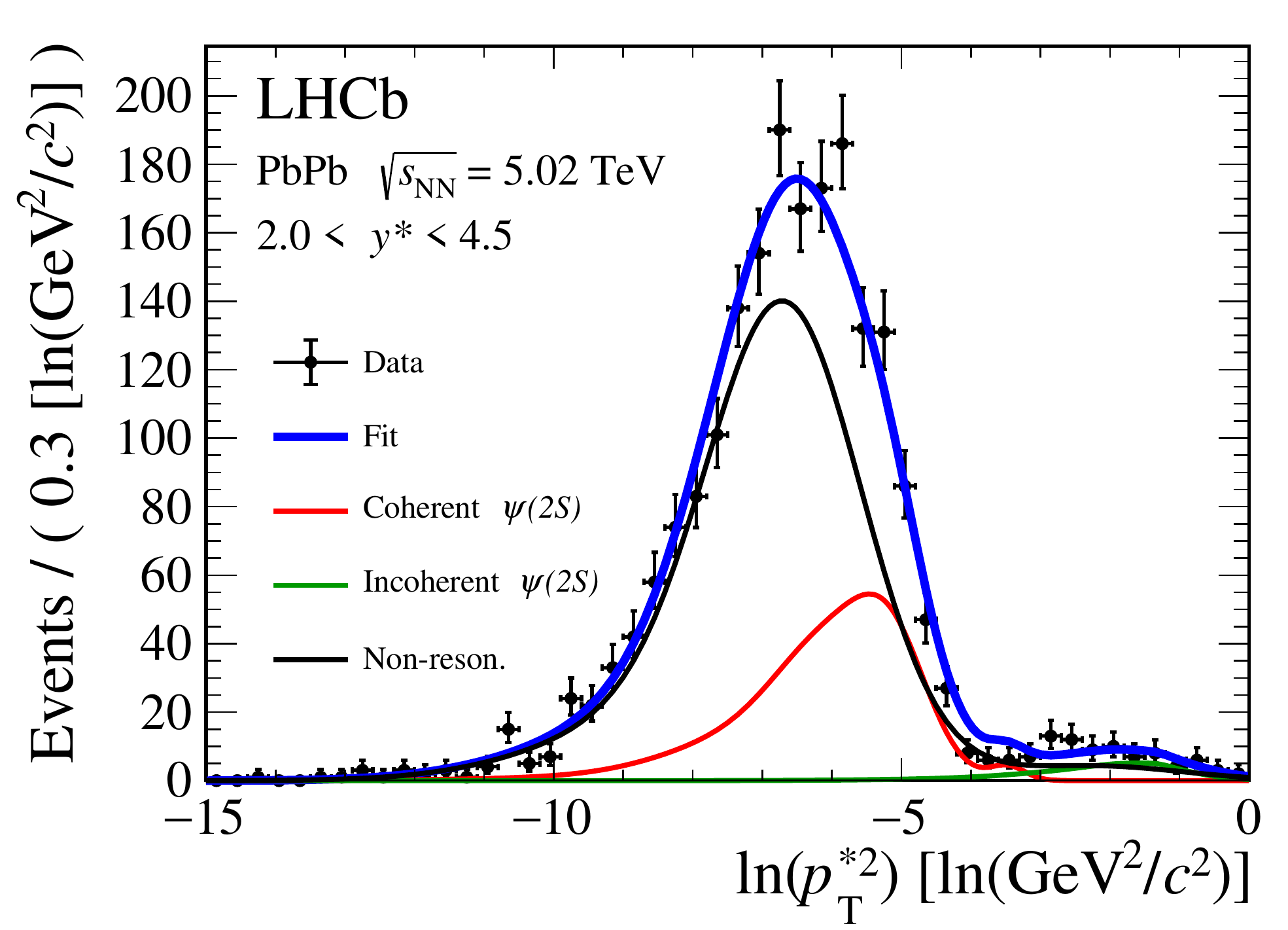}
        \put(-45,100){\psitwos}
    \end{minipage}
    \hfil
    \caption{Fit to the invariant mass distribution of dimuon candidates (up) and the \logpt2 distribution fit of dimuon candidates within the $2.0<y^{*}<4.5$ range for \jpsi candidates (left) and
    \psitwos candidates (right), where the starred notation indicates that the observable is defined in the nucleus-nucleus centre-of-mass frame.}
    \label{fig:2d}
\end{figure}

The charmonia candidates are reconstructed through the $\jpsi\rightarrow\mumu$ and $\psitwos\rightarrow\mumu$ decay channels using a PbPb data sample corresponding to an integrated luminosity of $228\pm10\,\mathrm{\mu b}^{-1}$, which was collected by the \lhcb experiment in 2018. 
According to the characteristics of UPCs, only events with low activity could be kept in the selection.
After that, 
the signal extraction is done through two steps.
A fit on the dimuon invariant mass spectrum is needed to estimate the non-resonant background yields within the \jpsi and \psitwos mass windows, then the fits to \jpsi and \psitwos transverse momentum distributions of selected candidates are performed to determine the coherent production events from the inclusive charmonia yields, as shown in Fig.~\ref{fig:2d}. 

\begin{figure}[htp]
    \centering
   \begin{center}
     \begin{minipage}[t]{0.41\linewidth}
        \centering
        \includegraphics[width=\linewidth]{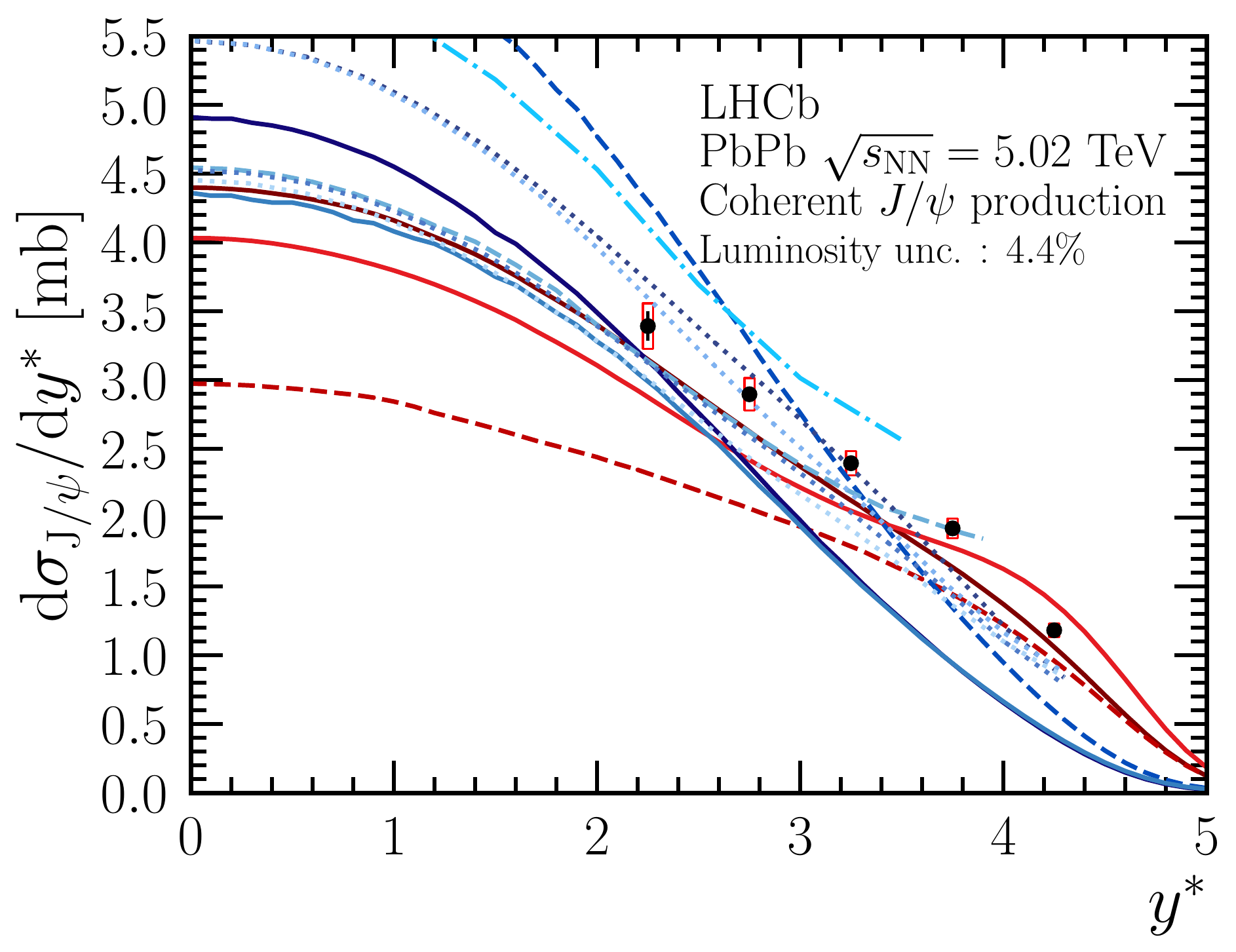}
     \end{minipage}  
      \begin{minipage}[t]{0.52\linewidth}
       \centering
    \includegraphics[width=\linewidth]{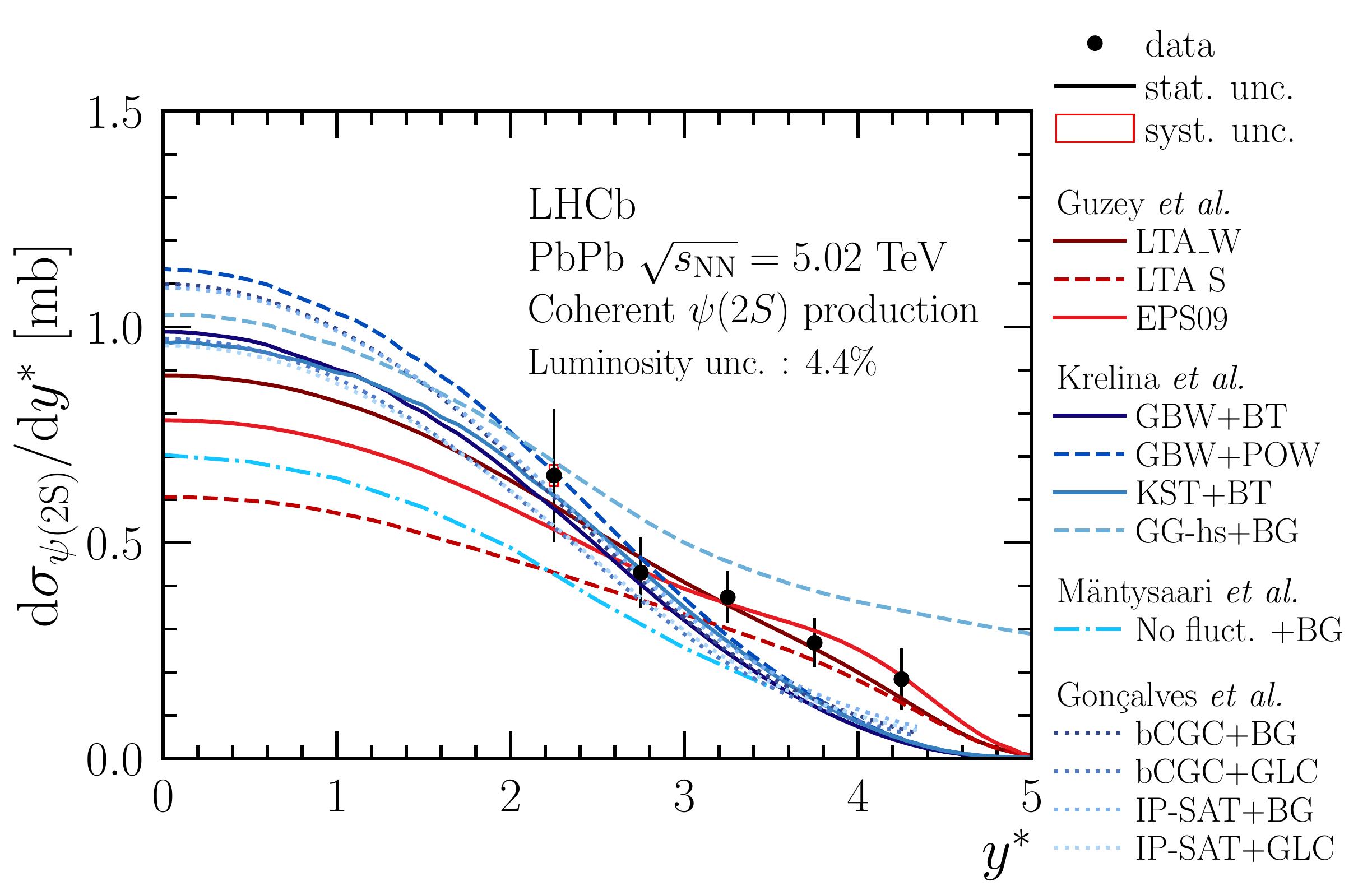}
     \end{minipage}
   \end{center}

     \begin{center}
        \begin{minipage}[t]{0.4\linewidth}
        \centering
       \includegraphics[width=\linewidth]{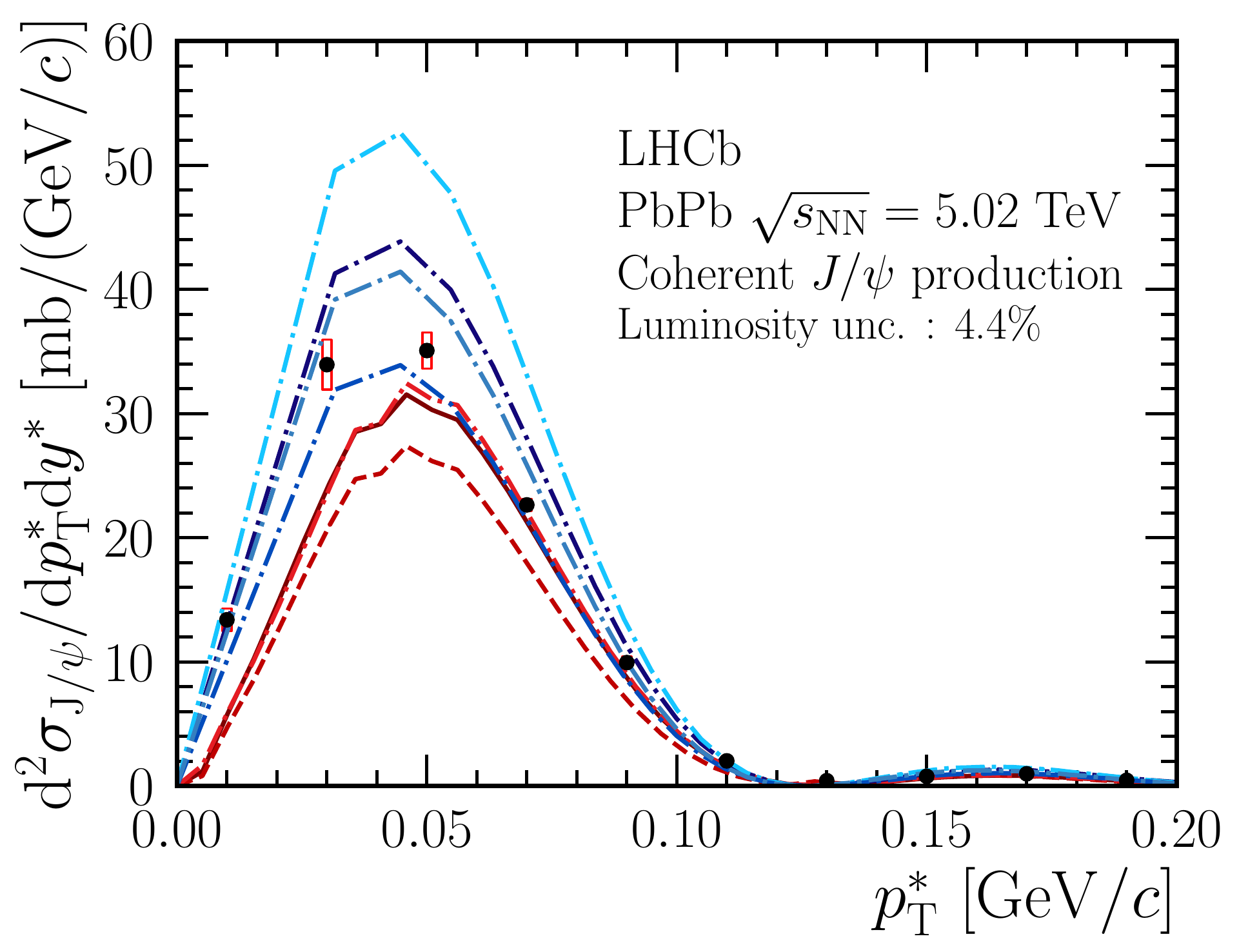}
      \end{minipage}
       \begin{minipage}[t]{0.52\linewidth}
       \centering
   \includegraphics[width=\linewidth]{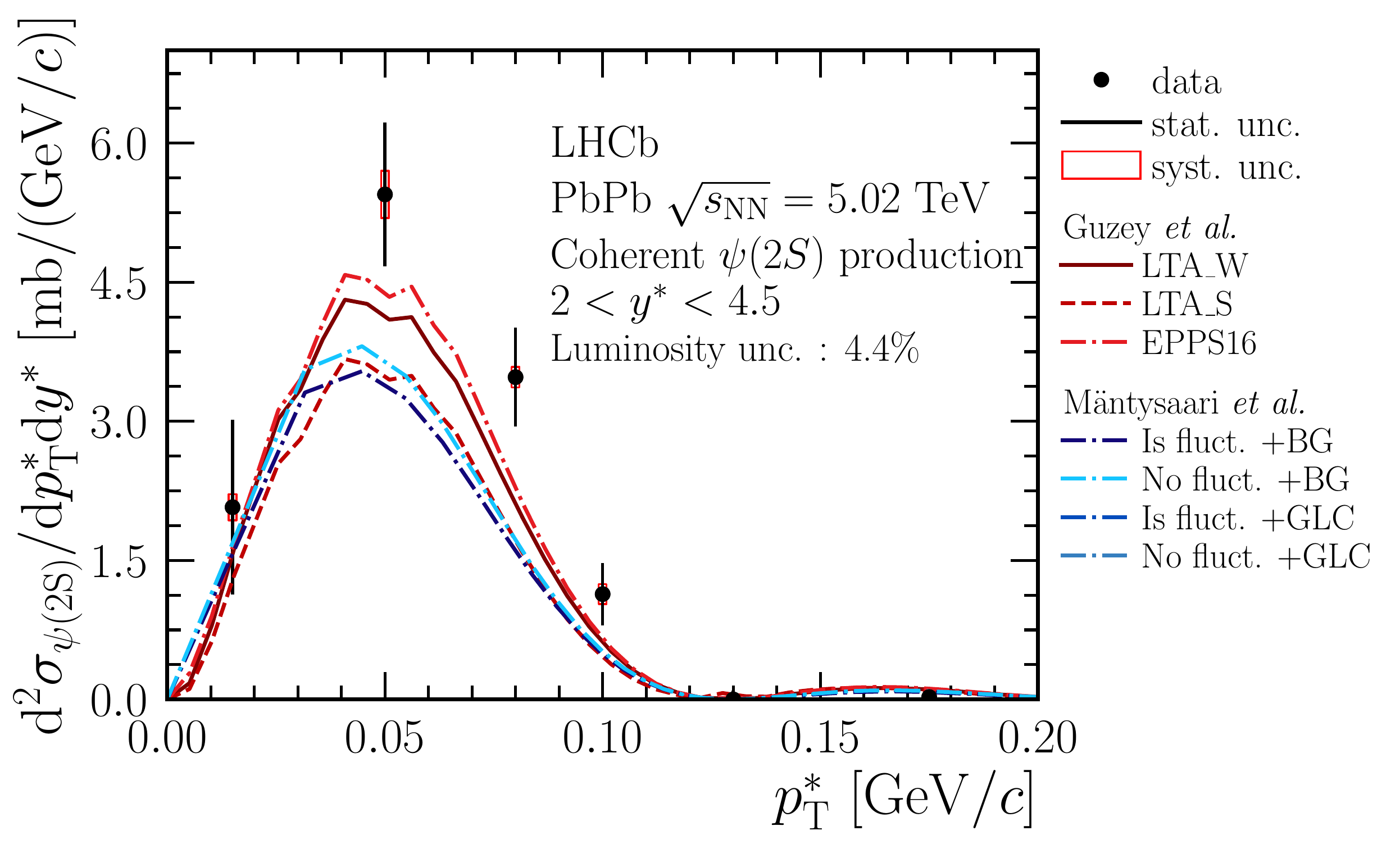}
      \end{minipage}      
     \end{center}
     \hfil
    \caption{Differential cross-section as a function $y^{*}$ (up) and $\pt^{*}$ (down) for coherent $\jpsi$ (left) and \psitwos (right) production. Compared to theoretical predictions, which are grouped as perturbative-QCD calculations (red lines) and colour-glass-condensate models (blue lines).}
    \label{fig:theo_y}
\end{figure}

\begin{figure}[htp]
\vspace{-1cm}
    \centering
    \begin{minipage}[t]{0.52\linewidth}
        \centering
        \includegraphics[width=\linewidth]{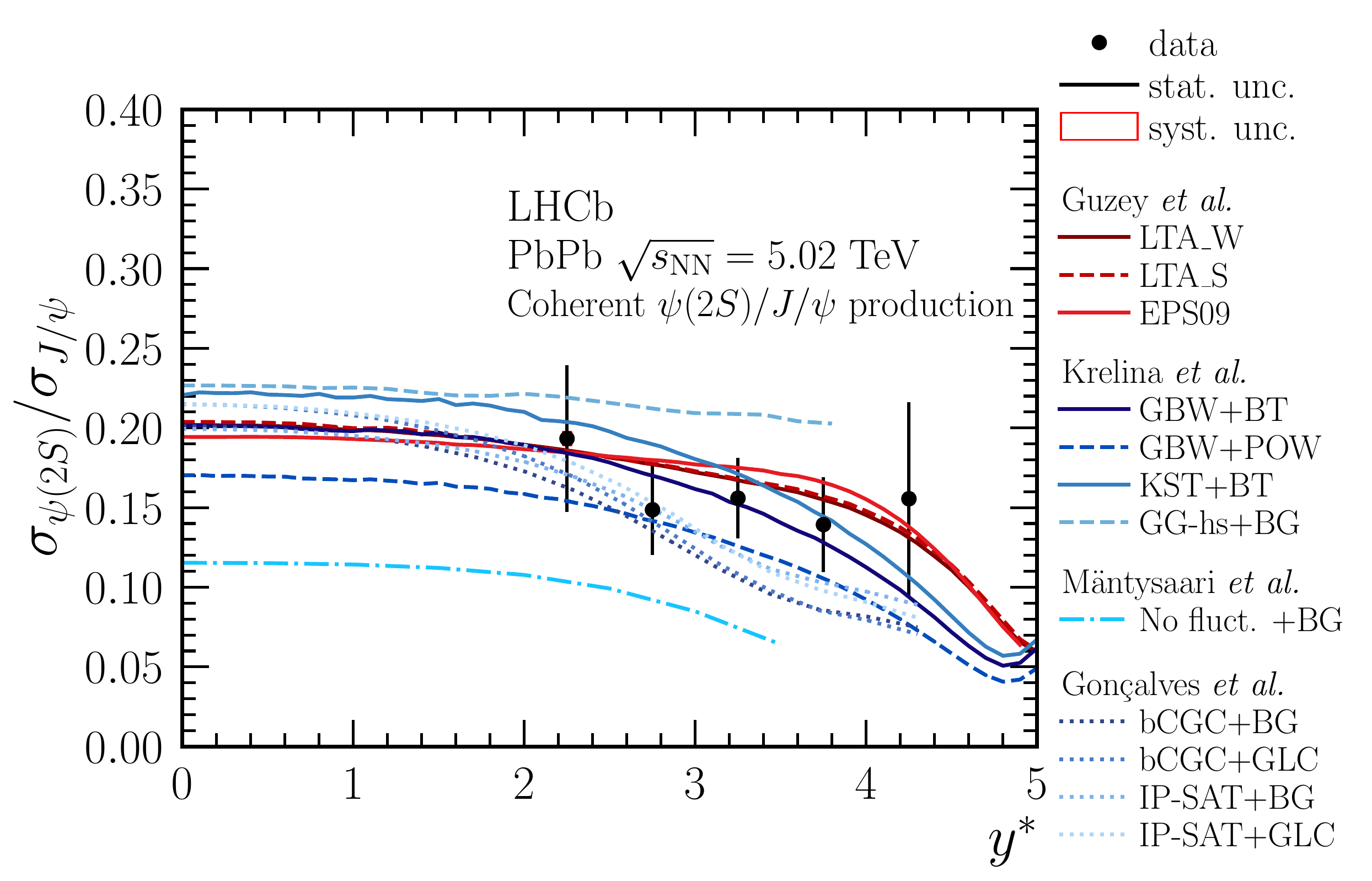}
    \end{minipage}
    \caption{Differential cross-section ratio of \psitwos to \jpsi as a function of $y^*$, compared to theoretical predictions, which are separated into perturbative QCD calculations (red lines) and colour-glass-condensate models (blue lines).}
    \label{fig:theo_ratio}
\end{figure}

The measured differential cross sections as a function of $y^{*}$ and $\pt^{*}$ of coherent \jpsi and \psitwos are shown in Fig.~\ref{fig:theo_y}, where the starred notation indicates that the observable is defined in the nucleus-nucleus centre-of-mass frame.
The differential cross-section ratio of coherent \psitwos to \jpsi production is calculated as a function of rapidity for the first time and shown in Fig.~\ref{fig:theo_ratio}.
Compared with theoretical predictions, which are grouped as perturbative-QCD calculations~\cite{Guzey_2016,2017access} and colour-glass-condensate (CGC) models~\cite{PhysRevC.84.011902,2018,Kopeliovich:2020has,PhysRevD.96.094027,Gon_alves_2005,20171,Mantysaari:2017dwh,2014arXiv1406.2877L}, it could be found that the measurements are in agreement with most of the prediction curves in general. 

\section{Study of \jpsi photo-production in PbPb peripheral collisions at $\sqrt{s_{NN}}= 5\tev$~\cite{LHCb:2021hoq}}
\begin{figure}[!htbp]
\vspace{-1cm}
    \centering
    \hfil
    \begin{minipage}[t]{0.49\linewidth}
        \centering
        \includegraphics[width=\linewidth]{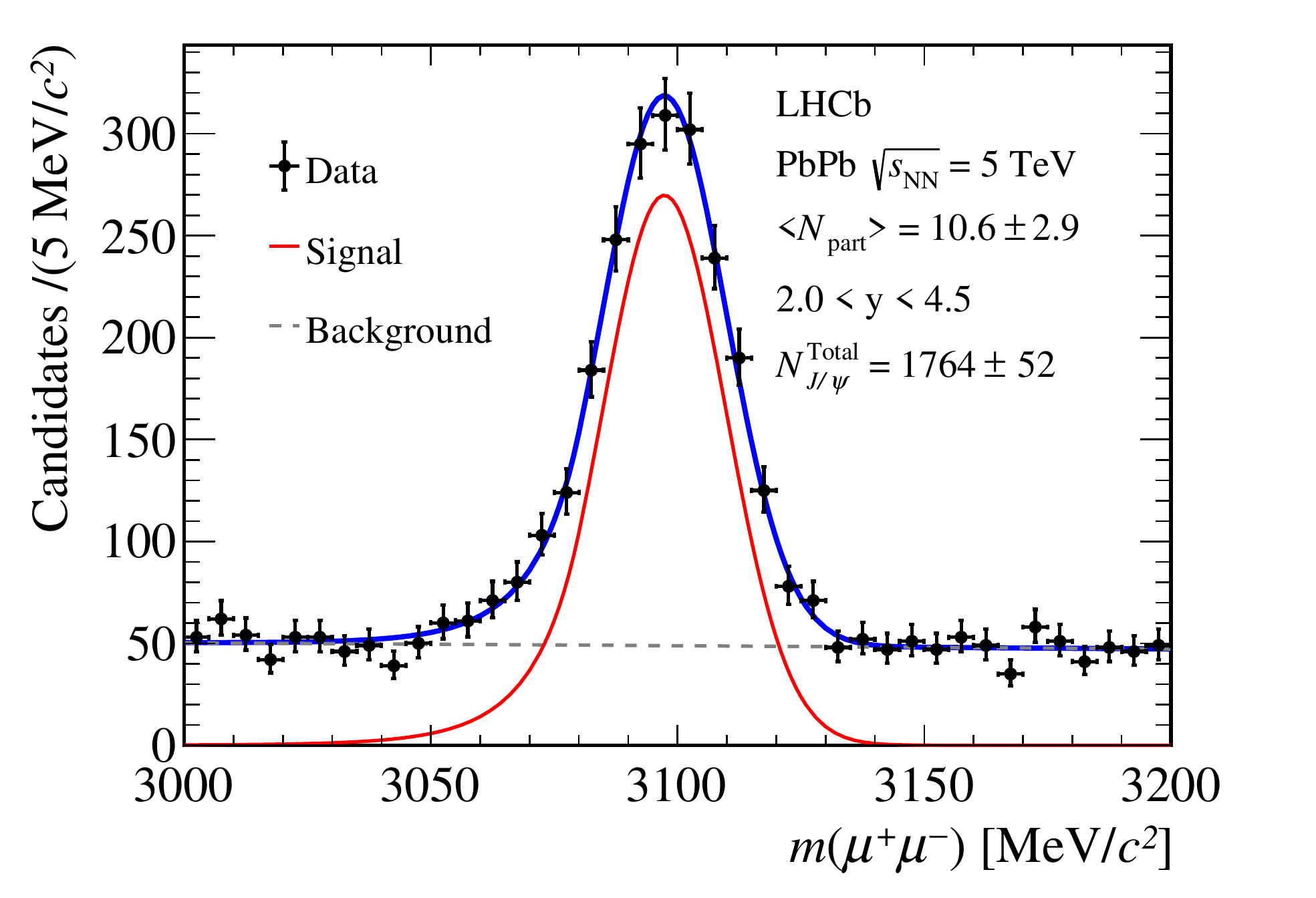}
    \end{minipage}
    \begin{minipage}[t]{0.49\linewidth}
        \centering
        \includegraphics[width=\linewidth]{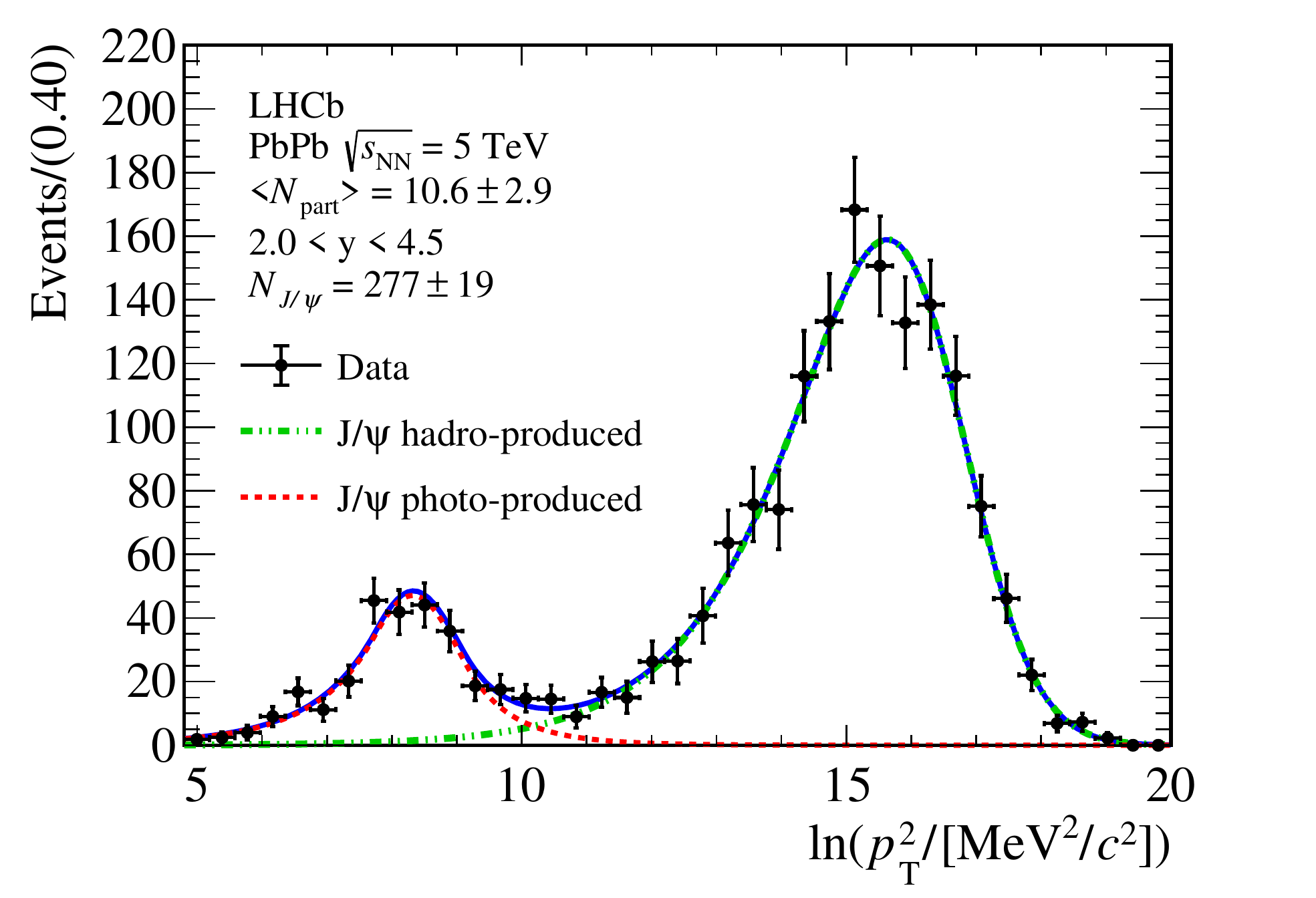}
    \end{minipage}
    \hfil
    \caption{Fit to the invariant mass spectrum of \jpsi candidates (left) and the \ensuremath{\ln(\pt^{2})}\xspace distribution fit for \jpsi candidates (right).}
    \label{PC2d}
\end{figure}

In peripheral collisions, the impact parameter is less than the sum of radii of the two colliding nuclei, so there is not only photo-nuclear interaction but also hadronic interaction. 
For hadronic productions, the gluon-gluon interaction produces a pair of charm quarks and lastly goes into the \jpsi meson. In this case, \jpsi meson will have transverse momentum typically between 1 and 2 \gevc, and for coherent photo-production, \jpsi would have a very small transverse momentum, less than 300 \mevc. 
Thanks to the high precision of the LHCb experiment, there are two visible clear peaks in the distribution of dimuon \logpt2 , which could be used to discriminate photo-produced and hadronically produced \jpsi, as shown in Fig.~\ref{PC2d}.

\begin{figure}[!htbp]
    \centering
    \begin{minipage}[t]{0.32\linewidth}
        \centering
        \includegraphics[width=\linewidth]{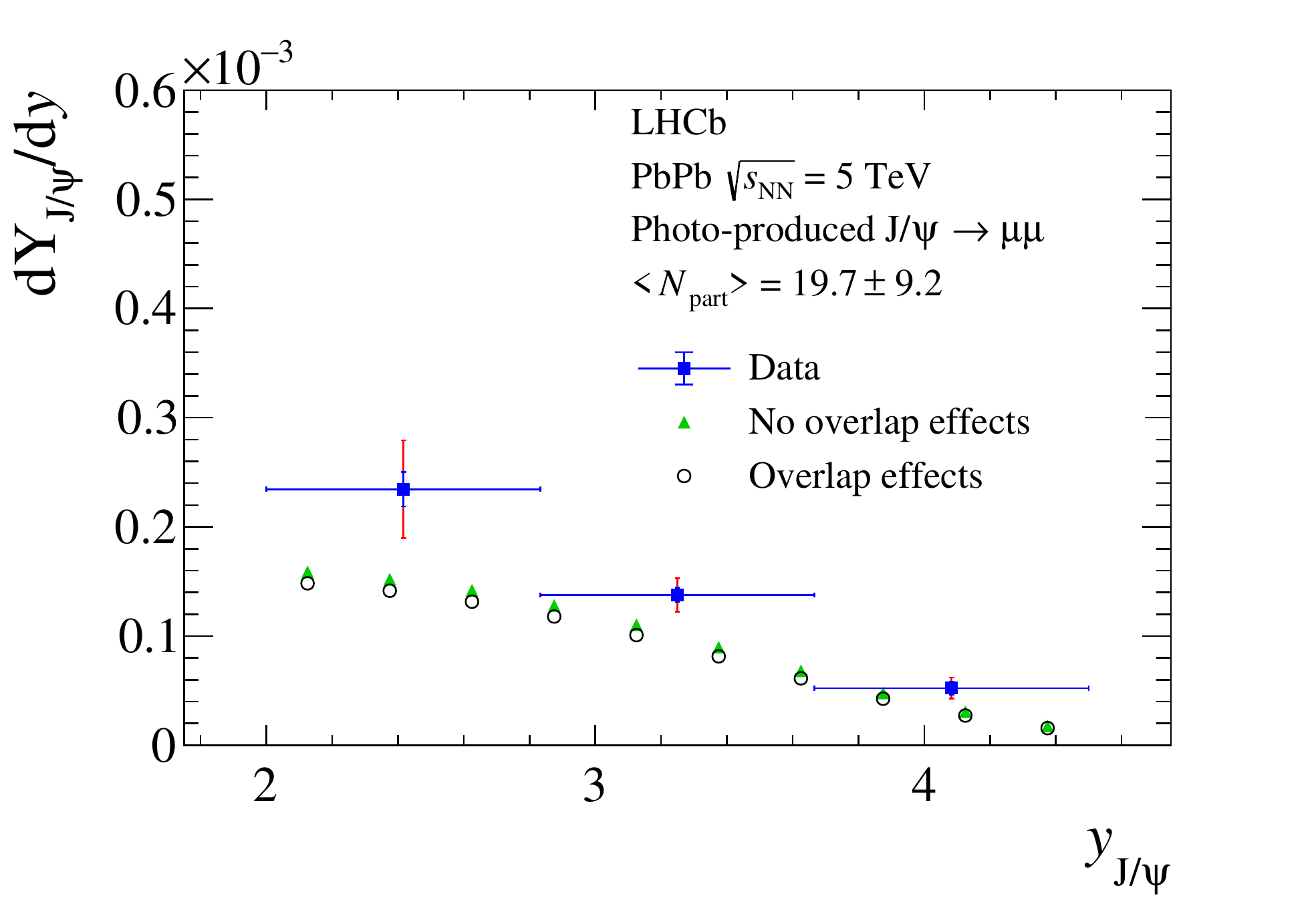}
    \end{minipage}
    \begin{minipage}[t]{0.32\linewidth}
        \centering
        \includegraphics[width=\linewidth]{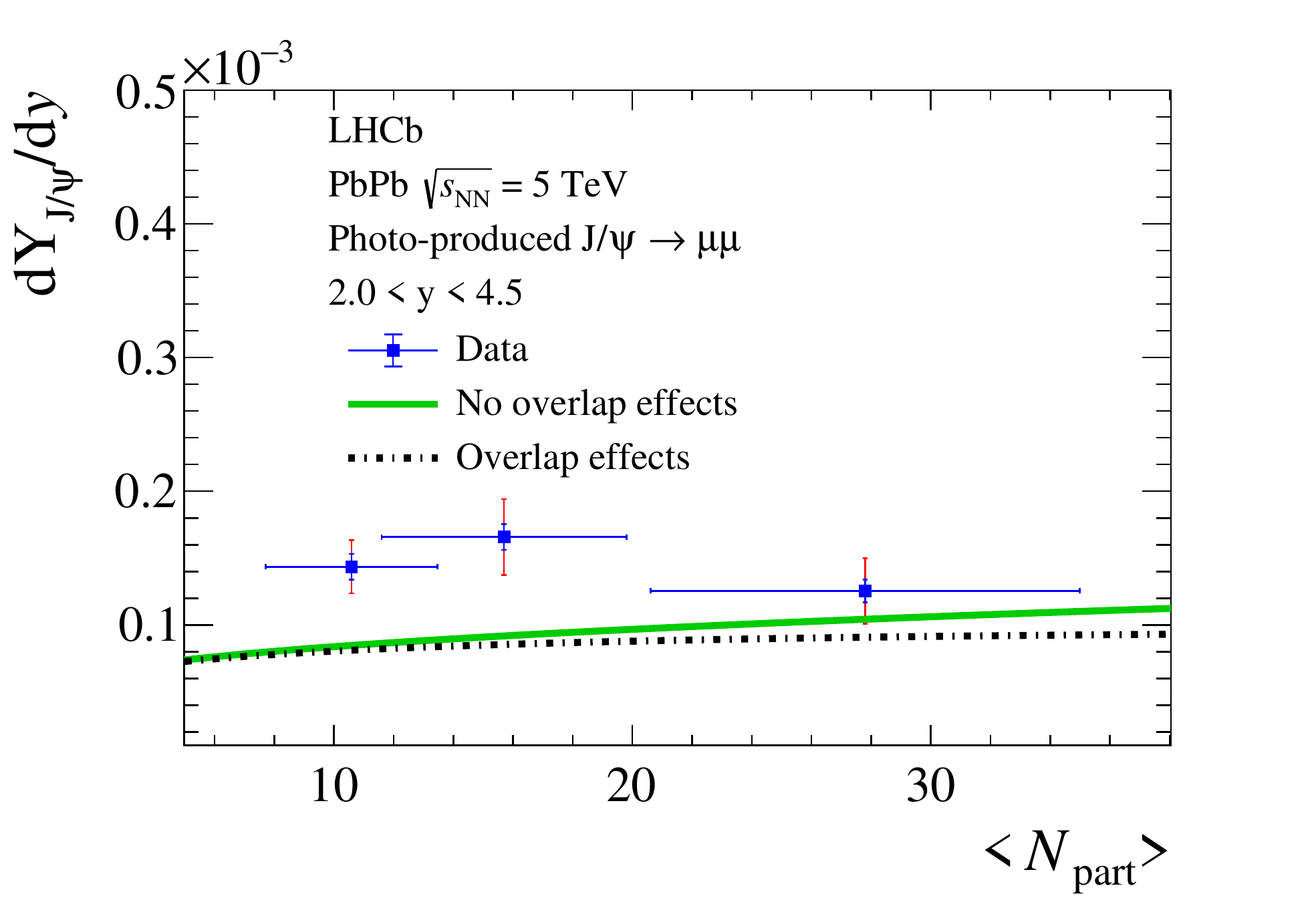}
    \end{minipage}
    \begin{minipage}[t]{0.32\linewidth}
        \centering
        \includegraphics[width=\linewidth]{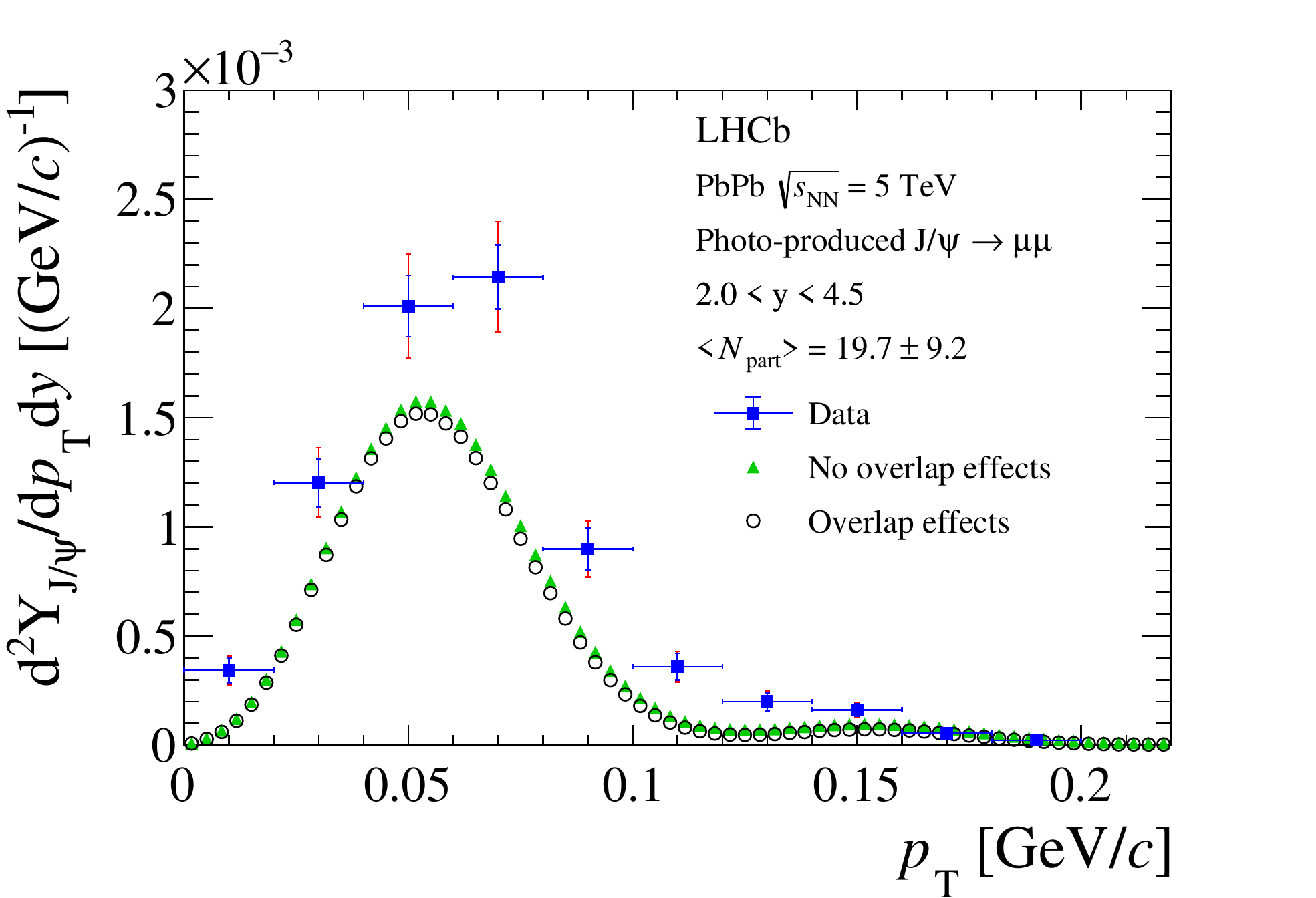}
    \end{minipage}
    \caption{The results of the differential photo-production yields of \jpsi versus the rapidity and the number of participants in collisions. Double-differential yields of the coherent \jpsi produced in peripheral PbPb collisions as a function of \pt (right).}
    \label{com}
\vspace{1cm}
\end{figure}

 The differential photo-production yields of \jpsi versus the rapidity, the number of participants in the collision, and the double-differential \jpsi photo-production yields versus transverse momentum as shown in Fig.~\ref{com}. 
For the phenomenological models, one scenario does not consider the destructive effect due to the overlap between the two nuclei, whereas the other takes it into account. 
In general, the trends of \jpsi photo-production measurements and theoretical predictions are consistent.
Meanwhile, the two theoretical curves do not show a significant difference because the collisions are peripheral and the nuclear overlapping effect is expected to increase in more central collisions. 
Currently, the measurement is limited to low values of $N_{part}$ due to detector limitation, 
and the measurement of \pt is consistent with the coherent \jpsi photo-production; the peak value of transverse momentum is around 60 \mevc.
\section{Conclusion}
We report the results of coherent  \jpsi, $\psi$(2S) production cross sections in UPCs, as well as their ratio.  
It is the first coherent \psitwos and \psitwos to \jpsi ratio measurement in forward rapidity region for UPC at \lhc, and the differential cross section of coherent \jpsi and \psitwos production in PbPb UPC is also measured as a function of $\pt^{*}$ for the first time. Also, it is the most precise measurement of \jpsi production in UPCs.
The production of photo-produced \jpsi mesons in peripheral PbPb collisions is measured and by far it is the most precise coherent photo-produced \jpsi by \lhcb. 

\addcontentsline{toc}{section}{References}
\bibliographystyle{LHCb}
\bibliography{sample,standard,LHCb-PAPER,LHCb-CONF,LHCb-DP,LHCb-TDR}
\end{document}

%% file: preamble.tex
\usepackage[top=1in, bottom=1.25in, left=1in, right=1in]{}
\usepackage{microtype}
\usepackage{lineno}  
\usepackage{xspace} 
\usepackage{caption} 

\usepackage{graphicx}  
\usepackage{color}
\usepackage{colortbl}
\graphicspath{{./figs/}} 
\DeclareGraphicsExtensions{.pdf,.PDF,png,.PNG}

\usepackage{amsmath} 
\usepackage{amssymb}
\usepackage{amsfonts}
\usepackage{upgreek} 

\newcommand*\patchAmsMathEnvironmentForLineno[1]{%
\expandafter\let\csname old#1\expandafter\endcsname\csname #1\endcsname
\expandafter\let\csname oldend#1\expandafter\endcsname\csname
end#1\endcsname
 \renewenvironment{#1}%
   {\linenomath\csname old#1\endcsname}%
   {\csname oldend#1\endcsname\endlinenomath}%
}
\newcommand*\patchBothAmsMathEnvironmentsForLineno[1]{%
  \patchAmsMathEnvironmentForLineno{#1}%
  \patchAmsMathEnvironmentForLineno{#1*}%
}
\AtBeginDocument{%
\patchBothAmsMathEnvironmentsForLineno{equation}%
\patchBothAmsMathEnvironmentsForLineno{align}%
\patchBothAmsMathEnvironmentsForLineno{flalign}%
\patchBothAmsMathEnvironmentsForLineno{alignat}%
\patchBothAmsMathEnvironmentsForLineno{gather}%
\patchBothAmsMathEnvironmentsForLineno{multline}%
\patchBothAmsMathEnvironmentsForLineno{eqnarray}%
}

\usepackage{hyperxmp}

\input{lhcb-symbols-def} 

\usepackage{cite} 
\usepackage{subfigure}
\usepackage{graphicx}
\usepackage{float}

%% file: lhcb-symbols-def.tex
\usepackage{xspace} 
\usepackage{upgreek}

\def\logpt2 {\ensuremath{\ln(\pt^{*2})}\xspace}   
\def\lhcb   {\mbox{LHCb}\xspace}

\def\lhc    {\mbox{LHC}\xspace}

\def\MagUp {\mbox{\em Mag\kern -0.05em Up}\xspace}

\ifthenelse{\boolean{uprightparticles}}%
{

 \def\Pmu         {\ensuremath{\upmu}\xspace}

 \def\Ppsi        {\ensuremath{\uppsi}\xspace}

 \def\PDelta      {\ensuremath{\Delta}\xspace}                 
 \def\PXi         {\ensuremath{\Xi}\xspace}                 
 \def\PLambda     {\ensuremath{\Lambda}\xspace}                 
 \def\PSigma      {\ensuremath{\Sigma}\xspace}                 
 \def\POmega      {\ensuremath{\Omega}\xspace}                 
 \def\PUpsilon    {\ensuremath{\Upsilon}\xspace}

 \def\PB      {\ensuremath{\mathrm{B}}\xspace}                 
                  
 \def\PD      {\ensuremath{\mathrm{D}}\xspace}

 \def\PJ      {\ensuremath{\mathrm{J}}\xspace}                 
 \def\PK      {\ensuremath{\mathrm{K}}\xspace}

 \def\Pi      {\ensuremath{\mathrm{i}}\xspace}

 \def\Ps      {\ensuremath{\mathrm{s}}\xspace}

 \def\thebaroffset{0.0em}
}
{

 \def\Pmu         {\ensuremath{\mu}\xspace}

 \def\Ppsi        {\ensuremath{\psi}\xspace}                 
                  
 \mathchardef\PDelta="7101
 \mathchardef\PXi="7104
 \mathchardef\PLambda="7103
 \mathchardef\PSigma="7106
 \mathchardef\POmega="710A
 \mathchardef\PUpsilon="7107
                  
 \def\PB      {\ensuremath{B}\xspace}                 
                  
 \def\PD      {\ensuremath{D}\xspace}

 \def\PJ      {\ensuremath{J}\xspace}                 
 \def\PK      {\ensuremath{K}\xspace}

 \def\Pi      {\ensuremath{i}\xspace}

 \def\Ps      {\ensuremath{s}\xspace}

 \def\thebaroffset{0.18em}
}
\newcommand{\offsetoverline}[2][\thebaroffset]{\kern #1\overline{\kern -#1 #2}}%

\makeatletter

\DeclareRobustCommand{\optbar}[1]{\shortstack{{\miniscule (\rule[.5ex]{1.25em}{.18mm})}
  \\ [-.7ex] $#1$}}

\def\mumu       {{\ensuremath{\Pmu^+\Pmu^-}}\xspace}

\def\squark    {{\ensuremath{\Ps}}\xspace}

\def\KorKbar {\kern \thebaroffset\optbar{\kern -\thebaroffset \PK}{}\xspace}

\def\DorDbar {\kern \thebaroffset\optbar{\kern -\thebaroffset \PD}\xspace}

\def\B       {{\ensuremath{\PB}}\xspace}

\def\BorBbar {\kern \thebaroffset\optbar{\kern -\thebaroffset \PB}\xspace}

\def\Bd      {{\ensuremath{\B^0}}\xspace}

\def\BdorBdbar {\kern \thebaroffset\optbar{\kern -\thebaroffset \Bd}\xspace}

\def\Bs      {{\ensuremath{\B^0_\squark}}\xspace}

\def\BsorBsbar {\kern \thebaroffset\optbar{\kern -\thebaroffset \Bs}\xspace}

\def\jpsi     {{\ensuremath{{\PJ\mskip -3mu/\mskip -2mu\Ppsi\mskip 2mu}}}\xspace}
\def\psitwos  {{\ensuremath{\Ppsi{(2S)}}}\xspace}

\def\Y#1S{\ensuremath{\PUpsilon{(#1S)}}\xspace}

\def\LorLbar     {\kern \thebaroffset\optbar{\kern -\thebaroffset \PLambda}\xspace}

\def\AT#1     {\ensuremath{A_{\mathrm{T}}^{#1}}\xspace}           

\def\C#1      {\ensuremath{\mathcal{C}_{#1}}\xspace}                       
\def\Cp#1     {\ensuremath{\mathcal{C}_{#1}^{'}}\xspace}                    
\def\Ceff#1   {\ensuremath{\mathcal{C}_{#1}^{\mathrm{(eff)}}}\xspace}        
\def\Cpeff#1  {\ensuremath{\mathcal{C}_{#1}^{'\mathrm{(eff)}}}\xspace}       
\def\Ope#1    {\ensuremath{\mathcal{O}_{#1}}\xspace}                       
\def\Opep#1   {\ensuremath{\mathcal{O}_{#1}^{'}}\xspace}                    

\newcommand{\aunit}[1]{\ensuremath{\text{\,#1}}}       

\newcommand{\tev}{\aunit{Te\kern -0.1em V}\xspace}
\newcommand{\gev}{\aunit{Ge\kern -0.1em V}\xspace}
\newcommand{\mev}{\aunit{Me\kern -0.1em V}\xspace}
\newcommand{\kev}{\aunit{ke\kern -0.1em V}\xspace}
\newcommand{\ev}{\aunit{e\kern -0.1em V}\xspace}
\newcommand{\mevc}{\ensuremath{\aunit{Me\kern -0.1em V\!/}c}\xspace}
\newcommand{\gevc}{\ensuremath{\aunit{Ge\kern -0.1em V\!/}c}\xspace}
\newcommand{\mevcc}{\ensuremath{\aunit{Me\kern -0.1em V\!/}c^2}\xspace}
\newcommand{\gevcc}{\ensuremath{\aunit{Ge\kern -0.1em V\!/}c^2}\xspace}

\def\gsim{{~\raise.15em\hbox{$>$}\kern-.85em
          \lower.35em\hbox{$\sim$}~}\xspace}
\def\lsim{{~\raise.15em\hbox{$<$}\kern-.85em
          \lower.35em\hbox{$\sim$}~}\xspace}

\def\pt         {\ensuremath{p_{\mathrm{T}}}\xspace}

\def\tell1  {TELL1\xspace}
\def\ukl1   {UKL1\xspace}

